\documentclass[12pt]{article}
\usepackage{epsfig,amssymb,amsmath,psfrag}

\def \bl  {\begin{align*}}
\def \el  {\end{align*}}

\def \be  {\begin{equation}}
\def \ee  {\end{equation}}
\def \ba  {\begin{eqnarray}}
\def \ea  {\end{eqnarray}}
\def \baa {\begin{eqnarray*}}
\def \eaa {\end{eqnarray*}}
\def \bb  {\begin {thebibliography} }
\def \eb  {\end{thebibliography}}
\def \lab #1 {\label{#1}}

\newcommand{\beq}{\begin{equation}}
\newcommand{\eeq}{\end{equation}}
\newcommand{\beqa}{\begin{eqnarray}}
\newcommand{\eeqa}{\end{eqnarray}}

\def\l<{\langle}
\def\r>{\rangle}

\newcommand \ov [1] {\overline{#1}}

\def\XXint#1#2#3{{\setbox0=\hbox{$#1{#2#3}{\int}$}
     \vcenter{\hbox{$#2#3$}}\kern-.5\wd0}}

\renewcommand{\title}[1]{\vbox{\center\LARGE{#1}}\vspace{5mm}}
\renewcommand{\author}[1]{\vbox{\center#1}\vspace{5mm}}

\begin{document}

\thispagestyle{empty}

\begin{flushright}
CERN-PH-TH/2010-275\\
HU-EP-10/81\\
\end{flushright}

\vskip2.2truecm
\begin{center}
\vskip 0.2truecm {\Large\bf
{\Large Yangian symmetry of light-like Wilson loops}
}\\
\vskip 1truecm
{\bf  
}

\vskip 0.4truecm

\begingroup\bf\large
J. M. Drummond,
\endgroup
\vspace{0mm}

\begingroup
\textit{PH-TH Division, CERN,\\
 Geneva, Switzerland 
 }\\
\par
\endgroup

\vspace{1.3cm}

\begingroup\bf\large
L. Ferro,
\endgroup
\vspace{0mm}

\begingroup
\textit{Institut f\"ur Physik, Humboldt-Universit\"at zu Berlin, \\
Newtonstra{\ss}e 15, D-12489 Berlin, Germany}\par
\endgroup

\vspace{0.7cm}

\begingroup\bf\large
E. Ragoucy,
\endgroup
\vspace{0mm}

\begingroup
\textit{LAPTH, Universit\'e de Savoie, CNRS\\
B.P. 110, F-74941 Annecy-le-Vieux Cedex, France}\par
\endgroup

\end{center}

\vskip 1truecm 
\centerline{\bf Abstract} 
We show that a certain class of light-like Wilson loops exhibits a Yangian symmetry at one loop, or equivalently, in an Abelian theory. The Wilson loops we discuss are equivalent to one-loop MHV amplitudes in $\mathcal{N}=4$ super Yang-Mills theory in a certain kinematical regime. The fact that we find a Yangian symmetry constraining their functional form can be thought of as the effect of the original conformal symmetry associated to the scattering amplitudes in the $\mathcal{N}=4$ theory.

\medskip

 \noindent

\newpage
\setcounter{page}{1}\setcounter{footnote}{0}

\section{Introduction}

Scattering amplitudes in gauge theories exhibit many surprising features hinting at an extraordinary simplicity that is not apparent in direct Feynman graph calculations. This is demonstrated at tree-level by the remarkable simplicity of the Parke-Taylor formula for maximally-helicity-violating amplitudes \cite{Parke:1986gb}. 
Such simplicity continues to all tree-level amplitudes if one employs the on-shell recursive BCFW relations \cite{Britto:2004ap,Britto:2005fq} to construct them from their known singularity structure.

The level of simplification is even greater when considering the maximally supersymmetric theory, $\mathcal{N}=4$ super Yang-Mills. In this case the recursive tree-level relations simplify \cite{ArkaniHamed:2008gz,Brandhuber:2008pf} and admit a closed-form solution \cite{Drummond:2008cr}. Furthermore the $\mathcal{N}=4$ theory exhibits a very large symmetry algebra. On the colour-ordered tree-level amplitudes the original superconformal symmetry of the Lagrangian combines with another copy of superconformal symmetry, called dual superconformal symmetry \cite{Drummond:2008vq} to form the Yangian of the superconformal algebra \cite{Drummond:2009fd}. The individual BCFW terms are each invariants under the full Yangian symmetry. They can be thought of as particular contour choices in the Grassmannian integral of \cite{ArkaniHamed:2009dn} (or equivalently its T-dual version \cite{Mason:2009qx,ArkaniHamed:2009vw}) which collects together all Yangian invariant objects into a single simple formula \cite{Drummond:2010qh,Drummond:2010uq,Korchemsky:2010ut}.

At loop level it has recently been realised that the above statements all hold at the level of the (unregulated) planar integrand. The integrand at a given loop order can be constructed from its singularities via a generalisation of the BCFW recursion relations and, remarkably, each term is individually invariant under the full Yangian symmetry up to a total derivative \cite{ArkaniHamed:2010kv}. At the level of the actual amplitudes the situation with the full symmetry is less clear, one issue being that the amplitudes are infrared divergent and thus require regularisation. A particularly useful regulator is the one obtained by introducing vacuum expectation values for the scalar fields \cite{Alday:2009zm,Henn:2010bk,Henn:2010ir}. This regulator preserves the dual conformal symmetry so that the resulting integrals are invariant. For work relating this picture to higher dimensions see \cite{Bern:2010qa,CaronHuot:2010rj,Dennen:2010dh}.
 
It has been known for a while that amplitudes in $\mathcal{N}=4$ super Yang-Mills are connected with light-like Wilson loops both at strong coupling (via the AdS/CFT correspondence) \cite{Alday:2007hr} and in perturbation theory \cite{Drummond:2007aua,Brandhuber:2007yx}. While at strong coupling the dependence on the helicity configuration of the amplitude appears as a subleading effect, in perturbation theory the correspondence with Wilson loops was originally limited only to the MHV amplitudes (with direct evidence of the correspondence coming up to two loops and six points \cite{Bern:2008ap,Drummond:2008aq}). However the amplitude/Wilson loop relation has recently been generalised to cover all helicity configurations \cite{Mason:2010yk,CaronHuot:2010ek}.

From the Wilson loop perspective the dual conformal symmetry of the scattering amplitudes is the natural conformal symmetry of the light-like Wilson loops. Its effects are taken into account via an anomalous Ward identity \cite{Drummond:2007cf,Drummond:2007au} which fixes the finite part of the Wilson loops (or equivalently MHV amplitudes) up to a function of conformally invariant cross-ratios. The Ward identity therefore expresses the consequence of the dual conformal symmetry of the scattering amplitudes. What is not clear is how the original conformal symmetry of the scattering amplitudes is realised beyond tree-level. The question of what happens to the original conformal symmetry at one loop has been addressed before in several papers. In particular the non-invariance of the one-loop amplitudes itself is not just due to the obvious breaking due to the presence of infrared divergences. A further effect can be traced to the holomorphic anomaly which gives a contact term variation even at tree level due to collinear singularities \cite{Bargheer:2009qu,Korchemsky:2009hm,Sever:2009aa,Beisert:2010gn}. Here, by appealing to Yangian structure of the underlying algebra we will be able to give a simple realisation of the symmetry on the one-loop amplitudes.

There are two ways of looking at the full Yangian symmetry. The first is to treat the original superconformal symmetry of the scattering amplitudes as fundamental. The additional dual conformal symmetry then extends this symmetry algebra to its Yangian \cite{Drummond:2009fd}. The second way exchanges the roles of the original and dual copies of the superconformal symmetry \cite{Drummond:2010qh}. The equivalence of these two pictures should be thought of as the algebraic realisation of the T-duality which maps scattering amplitudes to Wilson loops \cite{Ricci:2007eq,Berkovits:2008ic,Beisert:2008iq}.

The second way of thinking about the symmetry is more important in this paper. We will show that there is a natural (dual) conformally invariant finite quantity, described most naturally in terms of Wilson loops, which exhibits a Yangian symmetry. The finite quantity in question is the ratio of Wilson loops defined in \cite{Alday:2010ku,Gaiotto:2010fk}, corresponding to a choice of OPE channel when one considers expanding some subset of light-like edges around its totally collinear configuration.

We will be working with Wilson loops with special light-like contours contained in a two-dimensional subspace of the full spacetime \cite{Alday:2009yn}. The one-loop form of the light-like Wilson loops has been known for some time \cite{Brandhuber:2007yx} to be equivalent to the one-loop MHV amplitudes in $\mathcal{N}=4$ super Yang-Mills theory \cite{Bern:1994zx}. In the special two-dimensional kinematics they can be expressed purely in terms of logarithms \cite{Alday:2009yn,Heslop:2010kq}. 
Recently also two-loop functions have become available for six points in general kinematics \cite{DelDuca:2010zg,Goncharov:2010jf} and for an arbitrary number of points in the two-dimensional setup \cite{DelDuca:2010zp,Heslop:2010kq,Gaiotto:2010fk}.

In the two-dimensional kinematics the conformal symmetry of the Wilson loops is broken to an $sl(2)\oplus sl(2)$ subalgebra of the full conformal algebra $sl(4)$. The extra symmetry we find then corresponds to two commuting copies of the Yangian $Y(sl(2))$ and is best called a Yangian symmetry of the light-like Wilson loop. It should be thought of as the remaining effects of the original conformal symmetry of the scattering amplitudes in the special kinematics. 

We begin by discussing representations of Yangians in section 2. We will construct multi-parameter representations based on the coproduct and the freedom to change basis at each stage in building up the representation on a tensor product space. Then in section 3 we will construct some simple Yangian invariants. Of particular relevance is the fact that we find non-trivial logarithmic functions as possible invariants. In section 4 we discuss the geometrical setup of light-like Wilson loops in the restricted two-dimensional kinematics. We then go on to show that a natural finite, conformally invariant ratio constructed from the Wilson loops is actually invariant under two commuting copies of the Yangian $Y(sl(2))$. The quantity we find to be invariant is exactly the ratio defined in \cite{Alday:2010ku} corresponding to a particular choice of OPE channel for expanding the Wilson loops near a multi-collinear limit.

\section{Representations of Yangians}
\label{Yreps}

We are interested in particular kinds of representations of Yangian algebras \cite{Drinfeld:1985rx,Drinfeld:1987sy} based on the oscillator representation of the underlying algebra. We will consider $sl(m)$ as an example but the reasoning works also for $sl(m|n)$ if one includes both bosonic and fermionic oscillators.
So we will consider the representation  of $sl(m)$ given by
\be
J^{A}{}_{B} = W^{A} \frac{\partial}{\partial W^{B}} - \tfrac{1}{m} \delta^A_B W^C \frac{\partial}{\partial W^C}\,.
\ee
We prefer to write the oscillators as variables $W^A$ and derivatives $\partial/\partial W^B$ for $A,B=1,\ldots,m$ since we will eventually be interested in the space of invariant functions. The operator
\be
h = W^C\frac{\partial}{\partial W^C}
\ee
is central and so we can decompose the space of functions of the $W^A$ into those of fixed degrees of homogeneity $h$. Thus we can think of $W$ as homogeneous coordinates on $\mathbb{C}\mathbb{P}^{m-1}$. Our representation acts on functions with fixed degrees of homogeneity on this space (which we will denote as $\mathcal{F}(\mathbb{CP}^{m-1})$). In practice we will be interested in the case $h=0$.

To obtain a representation of the Yangian of the algebra we can apply the evaluation map \cite{Chari:1991xx} which constructs the level-one operator $J^{(1)}$ in terms of the algebra generators $J$. We could do this explicitly but there is a shortcut to the answer. In order to represent the level-one operator $J^{(1)}{}^{A}{}_{B}$ we need to write down an operator in the adjoint representation. Since the operator $h$ is central (and so can be assigned some fixed numerical value) our only choice is 
\be
J_\nu^{(1)}{}^{A}{}_{B} = \nu \biggl( W^A \frac{\partial}{\partial W^B} - \tfrac{1}{m}\delta^A_B W^C \frac{\partial}{\partial W^C} \biggr) \,.
\ee
The parameter $\nu$ is free. Together the operators $J^{A}{}_{B}$ and $J_\nu^{(1)}{}^{A}{}_{B}$ generate the Yangian $Y(sl(m))$. More precisely we have defined a representation $\pi_\nu$ of the Yangian which depends on a parameter $\nu$. The representation takes the form
\begin{align}
\pi_\nu \bigl(J^{A}{}_{B}) &= W^{A} \frac{\partial}{\partial W^{B}} - \tfrac{1}{m} \delta^A_B W^C \frac{\partial}{\partial W^C}\,, \\
\pi_\nu \bigl(J^{(1)}{}^{A}{}_{B}) &= \nu \biggl( W^{A} \frac{\partial}{\partial W^{B}} - \tfrac{1}{m} \delta^A_B W^C \frac{\partial}{\partial W^C}\, \biggr).
\end{align}

The Yangian is a Hopf algebra so we can construct further representations acting on the tensor product $\mathcal{F}(\mathbb{CP}^{m-1}) \otimes \ldots \otimes \mathcal{F}(\mathbb{CP}^{m-1})$ by using the coproduct\footnote{Here we use $a,b,c$ to denote adjoint indices.}
\begin{align}
\Delta J_a &= J_a \otimes 1 + 1 \otimes J_a\,,\\
\Delta J^{(1)}_a &= J^{(1)}_a \otimes 1 + 1 \otimes J^{(1)}_a + f_{a}{}^{cb} J_b \otimes J_c\,.
\end{align}
We can project the RHS of each of these relations with $\pi_{\nu_1} \otimes \pi_{\nu_2}$ to obtain a two-parameter representation acting on two sites,
\begin{align}
\pi_{\nu_1,\nu_2}(J^{A}{}_{B}) &= \sum_{i=1}^2 \biggl(W_i^A \frac{\partial}{\partial W_i^B} - \tfrac{1}{m} \delta^A_B W_i^C \frac{\partial}{\partial W_i^C} \biggr)\,,\\
\pi_{\nu_1,\nu_2}(J^{(1)}{}^{A}{}_{B}) &= \biggl(W_1^A \frac{\partial}{\partial W_1^C} W_2^C \frac{\partial}{\partial W_2^B} - (1,2)\biggr) + \sum_{i=1}^2 \nu_i \biggl(W_i^A \frac{\partial}{\partial W_i^B} -\tfrac{1}{m} \delta^A_B h_i\biggr)\,.
\end{align}
The operators
\be
h_i = \ W_i^C \frac{\partial}{\partial W_i^C} 
\ee
are central and so we can decompose the space of functions of the $W_i$ into spaces of fixed homogeneity in each of the $W_i$ separately.

One can continue and repeated application of the coproduct and projection with $\pi_{\nu_i}$ on the $i$th site yields the representation
\begin{align}
\pi_{\vec{\nu}}(J^{A}{}_{B}) &= \sum_{i=1}^n \biggl(W_i^A \frac{\partial}{\partial W_i^B} - \tfrac{1}{m}  \delta^A_B h_i\biggr)\,, \\
\pi_{\vec{\nu}}(J^{(1)}{}^{A}{}_{B}) &= \sum_{i<j} \biggl(W_i^A \frac{\partial}{\partial W_i^C}W_j^C \frac{\partial}{\partial W_j^B} - (i,j)\biggr) + \sum_{i=1}^n \nu_i \biggl(W_i^A \frac{\partial}{\partial W_i^B} -\tfrac{1}{m} \delta^A_B h_i\biggr)\,.
\end{align}
where $\vec{\nu} = (\nu_1,\ldots,\nu_n)$. We will mostly use the symbols $J^{A}{}_{B}$ and $J^{(1)}_{\vec{\nu}}{}^{A}{}_{B}$ to denote this representation of the level-zero and level-one generators. Note that the construction of the representation has provided an ordering of the sites from $1$ to $n$.

\section{Invariants}
\label{Yinvs}

Now let us consider functions of the $W_i$ which are invariant under the action of the Yangian generators $J$ and $J^{(1)}$. Firstly we make a general remark that when considering invariants we are free to add any amount of $J^{A}{}_{B}$ to $J_{\vec{\nu}}^{(1)}{}^{A}{}_{B}$ without changing the problem. We can use this freedom to set one of the $\nu_i$ to some fixed value, e.g. we could set $\nu_n =0$ if we wish.

If we are just interested in $sl(m)$-invariant functions of the $W_i$ then we can have any function of the invariant quantities,
\be
(i_1\ldots i_m) = W_{i_1}^{A_1}\ldots W_{i_m}^{A_m} \epsilon_{A_1\ldots A_m}\,.
\label{slninvariants}
\ee 
There are obviously no such quantities if we have fewer than $m$ sites as the above invariants are totally antisymmetric in all labels $i_1,\ldots i_m$. 

If we also require homogeneous functions with degree zero in all of the $W_i$ then we must consider functions of homogeneous ratios of the invariants in equation (\ref{slninvariants}). Let us consider $sl(2)$ as it is the simplest example and the one most relevant for this paper. The first possibility to form a homogeneous ratio is at four sites where we can write
\be
u=\frac{(13)(24)}{(14)(23)}\,.
\ee
This is the only independent invariant we can write. The only other possibility is related to $u$ using the cyclic identity $(ab)W_c^A + (bc)W_a^A + (ca)W_b^A =0$,
\be
\frac{(12)(34)}{(41)(23)} = 1-u\,.
\ee
Thus the $sl(2)$ invariant functions on four copies of $(\mathbb{CP}^1)$ are functions of $u$. Requiring that they are also Yangian invariant functions means we have to solve the equations
\be
J_{\vec{\nu}}^{(1)}{}^{A}{}_{B} f(u) = 0\,.
\ee
There are three independent equations here as the generators $J^{(1)}_{\vec{\nu}}{}^{A}{}_{B}$ are traceless. Obviously a constant function is always a solution of the equations.
We find that there is a non-trivial solution only if $\nu_1-\nu_3=\nu_2-\nu_4=2$. As we have discussed, although there are four $\nu_i$ our problem only really depends on three of them, or equivalently on the three independent differences. Since we have found two constraints from the condition of invariance there remains a one-parameter family of non-trivial invariant functions. They are given by hypergeometric functions,
\be
f_{\mu}(u) = \frac{(1-u)^{1+\mu}}{1+\mu}\,\, {}_2F_1(1,1+\mu,2+\mu;1-u)\,,\qquad  \mu=\tfrac{1}{2}(\nu_2-\nu_1)\,.
\ee
These functions represent the only homogeneous functions at four sites which are also Yangian invariants. Note that the representation of the Yangian was also constrained by the analysis. Of the three independent $\nu_i$, two were fixed. A particularly simple case is when we take $\mu=0$, in which case we have
\be
f_0(u) = (1-u) \,\,{}_2F_1(1,1,2;1-u) = \log u\,.
\ee
We will label this logarithmic invariant by 
\be
\log u = \log \frac{(13)(24)}{(14)(23)}=L(1,2,3,4)\,,
\ee 
to recall the order of the $W_i$ upon which it depends. Thus we have
\be
J^{A}{}_{B} L(1,2,3,4) = 0\,, \qquad J^{(1)}_{\vec{\nu}}{}^{A}{}_{B}L(1,2,3,4) = 0\,, \qquad \vec{\nu}=(1,1,-1,-1)\,.
\ee
Here we have used the freedom of shifting all the $\nu_i$ so that $\nu_1=1$.

A very simple way to obtain invariants for $n$ sites is simply to promote an invariant at $(n-1)$ sites. Suppose $Y_{n-1}(1,\ldots,n-1)$ is an invariant under the representation with labels $\vec{\mu}$ at $(n-1)$ sites. Then we can define an invariant at $n$ sites under the representation with labels $\vec{\nu}$ by the definition,
\be
Y_n(1,\ldots,n) \equiv Y_{n-1}(1,\ldots,n-1)\,.
\ee
Then since $Y(1,\ldots,n-1)$ is an invariant at $(n-1)$ sites we have
\be
J \,\, Y_{n-1}(1,\ldots,n-1)=0,\qquad J^{(1)}_{\vec{\mu}} \, Y_{n-1}(1,\ldots,n-1)=0\,.
\ee
It is then simple to see that $Y_n(1,\ldots,n)$ is an invariant at $n$ sites,
\be
J \,\, Y_n(1,\ldots,n)=0,\qquad J^{(1)}_{\vec{\nu}} \, Y_n(1,\ldots,n)=0\,,
\ee 
provided we choose the vector $\vec{\nu} = (\vec{\mu},\nu_n)$ for any value of $\nu_n$. This is exactly the adding operation of \cite{ArkaniHamed:2010kv}, taking into account the labels $\vec{\nu}$ defining the representation. Note that we could have introduced the site anywhere along the chain, i.e. we could have defined
\be
Y_n(1,\ldots,n) = Y_{n-1}(1,\ldots,i,i+2,\ldots,n)\,.
\ee
As an example we can consider five sites and construct invariants from the logarithmic invariant $L$; the combination
\be
a L(1,2,4,5) + b L(1,3,4,5) + c L(2,3,4,5)
\label{5ptfrom4pt}
\ee
is invariant provided we choose $\vec{\nu}=(1,1,1,-1,-1)$.

\section{Light-like Wilson loops}
We will consider Wilson loops defined on polygonal light-like contours in four-dimensional gauge theory. Our motivation is to understand how the integrable nature of planar $\mathcal{N}=4$ super Yang-Mills theory manifests itself in the form of such Wilson loops.

Wilson loops with cusps have ultra-violet divergences. We will write the polygonal light-like Wilson loops as follows,
\be
\log W_n = \sum_i [\text{UV div}]_i + F_n^{\rm anom}(x_1,\ldots,x_n) + I_n(u_1,\ldots,u_m)\,.
\ee
Here we have a specific divergences coming from each cusp denoted by $[\text{UV div}]_i$. The finite part has been split into two parts, $F_n^{\rm anom}$ and $I_n$.
The first is a contribution which satisfies the anomalous Ward identity due to conformal symmetry \cite{Drummond:2007cf,Drummond:2007au},
\be
K^\mu F_n^{\rm anom}(x_1,\ldots,x_n) = \Gamma_{\rm cusp}(\lambda) \sum_i (2 x_i^\mu - x_{i-1}^\mu - x_{i+1}^\mu)\log x_{i-1,i+1}^2\,.
\ee
This part could be taken to be the one-loop result multiplied by the cusp anomalous dimension. In this case it coincides with BDS ansatz part of the MHV scattering amplitude \cite{Bern:2005iz}. However the definition of $F_n^{\rm anom}$ is ambiguous because it can be modified by any function of the available conformal invariants $u_1,\ldots,u_k$ (here $k=3n-15$).

Depending on the choice of the definition of the anomalous part, there is an additional part which is just a function of conformal invariants, $I_n$. If we choose $F_n^{\rm anom}$ to coincide with the BDS ansatz for the MHV amplitude then $I_n$ is the standard definition of the `remainder function'. In this case it is non-zero only at two loops and beyond and for six or more points \cite{Drummond:2007bm,Drummond:2008aq,Bern:2008ap}. There are alternative definitions of the anomalous part which modify it by adding some function of invariants and subtracting the same function from the remainder function. An example is the definition of the `BDS-like' piece of \cite{Alday:2009yn} where the anomalous part depends only on the shortest distances $x_{i,i+2}^2$.

\begin{figure}
 \centerline{{\epsfysize4.5cm
\epsfbox{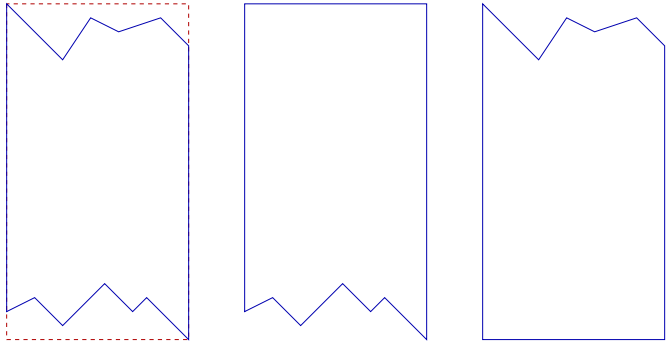}}}  \caption[]{\small The four different Wilson loops entering the definition of the ratio (\ref{ratio}). The reference square is shown by the dashed line. The bottom and top Wilson loops are obtained by replacing a sequence of edges by the corresponding part of the square.}
  \label{Wloops}
\end{figure}

A particularly interesting definition for the decomposition was made in \cite{Alday:2010ku}. In this case one picks two of the light-like edges and forms a light-like square by picking two more light-like lines intersecting them both. 
Then one can consider four different Wilson loops. The original Wilson loop, the Wilson loop on the square and the Wilson loops formed by replacing the top or bottom set of intermediate edges by the corresponding part of the square. This is best illustrated by Fig. \ref{Wloops}. 
One can define a conformally invariant quantity by the following ratio of the Wilson loops\footnote{In the generic situation there can be a single logarithmic divergence left in the ratio $r_n$. This will be absent in the ratio we study in this paper for two-dimensional loops. We would like to thank Johannes Henn for discussions on this point.},
\be
r_n = \log \biggl(\frac{W_n W_{\rm sq}}{W_{\rm top} W_{\rm bottom}}\biggr).
\label{ratio}
\ee
This quantity is a function of the cross-ratios $u_1,\ldots,u_{k}$.
Unlike the usual definition of the remainder function, $r_n$ is non-zero already at one loop. It is also not cyclic invariant since its definition required a choice of two special lines from which to form the square. This choice essentially corresponds to the choice of OPE `channel' in which one expands the Wilson loop over exchanged intermediate excited flux tube states \cite{Alday:2010ku}.

\begin{figure}
 \centerline{{\epsfysize4.5cm
\epsfbox{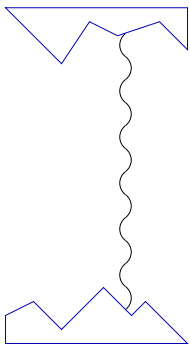}}}  \caption[]{\small An alternative picture for the one-loop diagrams contributing to $r_n$. The ratio of the Wilson loops defined in (\ref{ratio}) is equivalent to the connected diagrams in the correlator of the two loops shown here.}
  \label{upprlwr}
\end{figure}

The quantity $r_n$ is particularly simple at one loop. It corresponds to the connected part of the correlation between the two Wilson loops shown in Fig. \ref{upprlwr}. A further simplification is obtained when considering restricted two-dimensional kinematics as in \cite{Alday:2009yn}. In this case one needs an even number of sides to the Wilson loops, alternating in orientation between the $x^+$ direction and the $x^-$ direction as one travels round the loop. The number of independent cross-ratios is reduced in the two-dimensional kinematics. In fact there are $(n-6)$ independent ratios left from the original $(3n-15)$. Since $n$ is always even the first non-trivial ratio is therefore at eight points.

\begin{figure}
\psfrag{1}[cc][cc]{$x_1$}
\psfrag{2}[cc][cc]{\!$x_2$}
\psfrag{3}[cc][cc]{$x_3$}
\psfrag{i}[cc][cc]{$x_i$}
\psfrag{n}[cc][cc]{\,$x_n$}
\psfrag{ip}[cc][cc]{\!\!\!\!\!\!\!$x_{i+1}$}
\psfrag{w1}[cc][cc]{$w_1$}
\psfrag{w2}[cc][cc]{\,$\ov{w}_2$}
\psfrag{w3}[cc][cc]{\,\,\,$w_3$}
\psfrag{wi}[cc][cc]{\,\,$w_i$}
\psfrag{wn}[cc][cc]{\,\,$\ov{w}_n$}
 \centerline{{\epsfysize7.5cm
\epsfbox{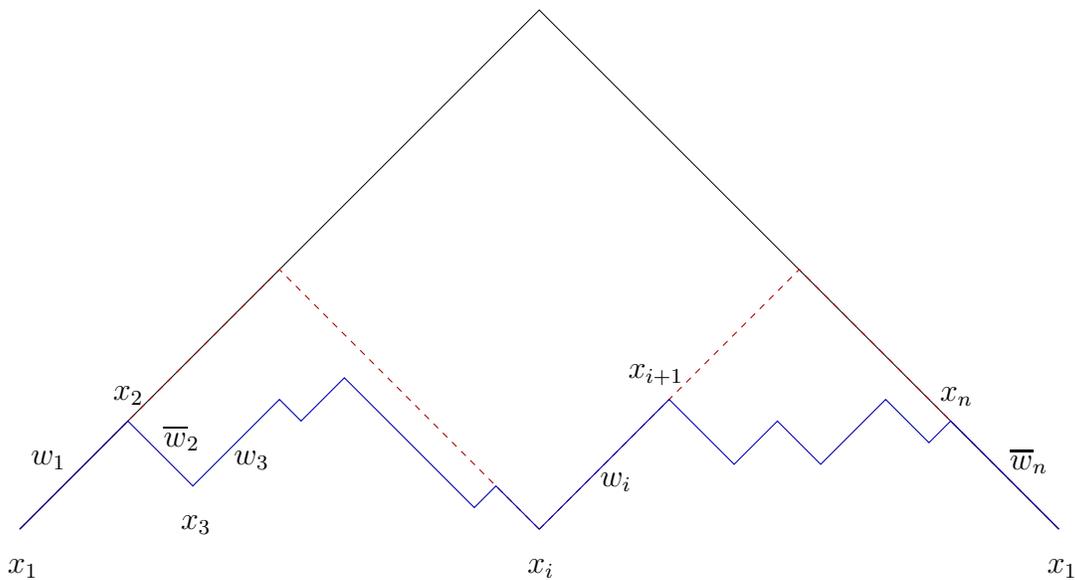}}}  \caption[]{\small The reference square denoted by the dashed line separates the edges of the Wilson loop into the two groups $\{1,\ldots,\ov{i-1}\}$ and $\{i,\ldots,\ov{n}\}$.}
  \label{2dpendiag}
\end{figure}

A very useful way to picture this is by drawing the Penrose diagram, putting two of the null sides of the loop at null infinity as in \cite{Alday:2009yn} (see Fig. \ref{2dpendiag}). Due to the light-like nature of the problem, it is very useful to describe the symmetry and kinematical dependence in terms of twistor variables. Here we mean twistor variables corresponding to light-like lines in the configuration space of the Wilson loop (corresponding to momentum twistors \cite{Hodges:2009hk} when viewed from the scattering amplitude perspective).

We recall that twistors can be defined from the position variables defining the light-like loop. Concretely we can write the light-like vectors defining the separations as a product of commuting spinors,
\be
x_{i}^{\alpha \dot\alpha} - x_{i+1}^{\alpha \dot\alpha} = \lambda_i^\alpha \tilde{\lambda}_i^{\dot\alpha}\,.
\ee
We can define the twistor variables $W_i^A=(\lambda_i^\alpha,\mu_i^{\dot\alpha})$ from the incidence relations,
\be
\mu_i^{\dot \alpha} = x_i^{\alpha \dot\alpha} \lambda_{i\alpha}\,.
\ee
The twistors $W_i^A$ transform linearly under $sl(4)$ conformal transformations, whose generators take the simple form,
\be
J^{A}{}_{B} = \sum_i \biggl( W_i^A \frac{\partial}{\partial W_i^B} - \tfrac{1}{4}\delta^A_B W_i^C\frac{\partial}{\partial W_i^C} \biggr)\,.
\ee
In the special two-dimensional kinematics the twistor variables are also restricted and preserve two commuting copies of $sl(2)$ inside $sl(4)$.
Specifically we can decompose the $W_i^A$ into upper and lower components each transforming under its own  $sl(2)$. The alternating orientations of the lines corresponds to an alternating between twistors transforming under the two copies of $sl(2)$. We take the odd-numbered twistors to transform under the first copy and the even-numbered ones to transform under the second copy,
\be
W_{2i+1} = 
\left(
\begin{matrix} w_{2i+1}\\0
\end{matrix}
\right),
\qquad
W_{2i} = 
\left(
\begin{matrix} 0 \\ \bar{w}_{2i}
\end{matrix}
\right)\,.
\ee
The generators of the two copies of $sl(2)$ are then
\be
J^{a}{}_{b} = \sum_{i \text{ odd}} \biggl( w_i^a \frac{\partial}{\partial w_i^b} - \tfrac{1}{2}\delta^a_b w_i^c\frac{\partial}{\partial w_i^c} \biggr)\,, \qquad \overline{J}^{\bar{a}}{}_{\bar{b}} =  \sum_{i \text{ even}} \biggl( \bar{w}_i^{\bar{a}} \frac{\partial}{\partial \bar{w}_i^{\bar{b}}} - \tfrac{1}{2}\delta^{\bar{a}}_{\bar{b}} \bar{w}_i^{\bar{c}}\frac{\partial}{\partial \bar{w}_i^{\bar{c}}} \biggr)\,.
\label{sl2s}
\ee
Here $a,b$ and $\bar{a},\bar{b}$ run from 1 to 2.

Referring back to Fig. \ref{2dpendiag} we can identify the twistor variables with the edges. This leads us to introduce some useful notation for $r_n$. We will write it as a function of the twistor variables which will be separated into two groups corresponding to the left and right groups of edges in Fig. \ref{2dpendiag}. Thus we have the general form
\be
r_{n}(1,\ov{2},\ldots,\ov{i-1}|i,\ov{i+1},\ldots,\ov{n})\,.
\ee
The vertical bar serves to indicate the separation into left and right groups, corresponding to the choice of OPE channel in \cite{Alday:2010ku}.

We recognise in (\ref{sl2s}) two copies of the representation of $sl(2)$ that we discussed previously. Thus we know how to extend each $sl(2)$ to its Yangian. We just take the additional generators $J_{\vec{\nu}}^{(1)}$ and $\overline{J}_{\vec{\bar{\nu}}}^{(1)}$ corresponding to the representations described in section \ref{Yreps}.
Let us consider the first non-trivial case $n=8$. We know from our previous analysis the form of the homogeneous invariants. Restricting to the simple integer weights we described in section \ref{Yinvs} we have invariants $L(1,3,5,7)$ and $L(\overline{2},\overline{4},\overline{6},\overline{8})$. The ratio $r_n$ is simple to compute since the relevant Wilson loops are known at one loop \cite{Brandhuber:2007yx} to coincide with the one-loop MHV amplitudes \cite{Bern:1994zx}. In the special two-dimensional kinematics all quantities can be expressed in terms of logarithms \cite{Alday:2009yn,Heslop:2010kq}.
Remarkably the function $r_8$ at one loop is none other than \cite{Gaiotto:2010fk}
\be
r_8(1,\ov{2},3,\ov{4}|5,\ov{6},7,\ov{8}) = g^2 L(1,3,5,7)L(\overline{2},\overline{4},\overline{6},\overline{8}) + \text{ const.}
\ee
It is therefore Yangian invariant under two copies of the Yangian $Y(sl(2))$ for the choices $\vec{\nu}=(1,1,-1,-1)$ and $\vec{\bar{\nu}} = (1,1,-1,-1)$ where the entries range over the odd and even values of $i$ respectively.

\begin{figure}
\psfrag{1}[cc][cc]{$x_1$}
\psfrag{2}[cc][cc]{\!\!\!\!\!\!\!$x_2$}
\psfrag{3}[cc][cc]{\,\,\,\,\,\,\,$x_3$}
\psfrag{i}[cc][cc]{$x_7$}
\psfrag{n}[cc][cc]{\,$x_{10}$}
\psfrag{xa}[cc][cc]{$x_a$}
\psfrag{xb}[cc][cc]{\,$x_{b}$}
\psfrag{ip}[cc][cc]{\!\!\!\!$x_{8}$}
\psfrag{w1}[cc][cc]{$w_1$}
\psfrag{w2}[cc][cc]{\,$\ov{w}_2$}
\psfrag{w3}[cc][cc]{\!\!\!\!\!\!$w_3$}
\psfrag{wi}[cc][cc]{\,\,$w_7$}
\psfrag{wn}[cc][cc]{\,\,$\ov{w}_{10}$}
 \centerline{{\epsfysize4.8cm
\epsfbox{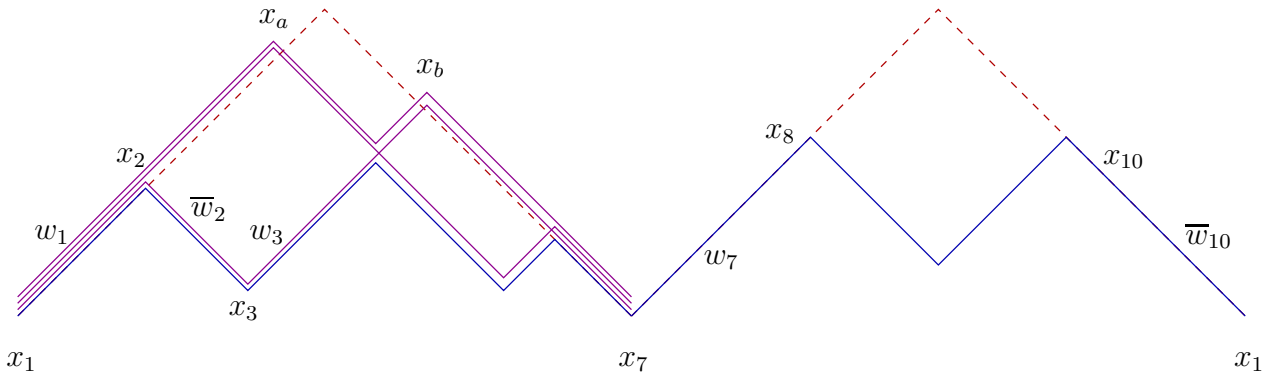}}}  \caption[]{\small The ten-sided Wilson loop can be expressed as the sum of the eight-sided Wilson loops passing through points $x_a$ and $x_b$ minus the one which passes through both.}
  \label{10ptto8pt}
\end{figure}

As pointed out in \cite{Gaiotto:2010fk} there is a very simple relation between $r_n$ and $r_{n-2}$ for a given choice of reference square. This amounts to the fact that at one loop the Wilson loops are additive in nature. As the simplest example one can write $r_{10}$ as a sum over three contributions of the form of $r_8$. We will label the three contributions with $a$, $b$ or $ab$ depending on whether the loops contain the additional points $x_a$, $x_b$ or both. The decomposition we have is
\be
r_{10} = r_{8,a} + r_{8,b} - r_{8,ab}\,.
\ee
A useful diagram to represent this is Fig. \ref{10ptto8pt}. 

\begin{figure}
\psfrag{w1}[cc][cc]{\small $1$}
\psfrag{w2}[cc][cc]{\small $\ov{2}$}
\psfrag{w3}[cc][cc]{\!\!\!\!\!\!\! \small $3$}
\psfrag{w4}[cc][cc]{\,\,\, \small $\ov{4}$}
\psfrag{w5}[cc][cc]{\small $5$}
\psfrag{w6}[cc][cc]{\small $\ov{6}$}
\psfrag{w7}[cc][cc]{\small $7$}
\psfrag{w8}[cc][cc]{\,\,\, \small $\ov{8}$}
\psfrag{w9}[cc][cc]{\!\!\!\!\!\!\! \small $9$}
\psfrag{w10}[cc][cc]{\small $\ov{10}$}
\psfrag{=}[cc][cc]{\small $=$}
\psfrag{+}[cc][cc]{\small $+$}
\psfrag{-}[cc][cc]{\small $-$}
 \centerline{{\epsfysize3.0cm
\epsfbox{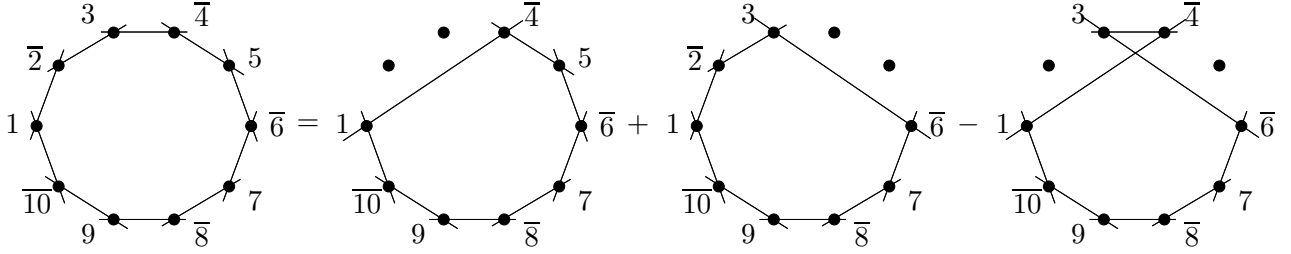}}}  \caption[]{\small The twistor space configurations of the three contributions to $r_{10}$. The points $x_a$ and $x_b$ correspond to the lines $(1\ov{4})$ and $(3\ov{6})$ respectively.}
  \label{twistor10ptto8pt}
\end{figure}

To see exactly how the functional dependence of $r_{10}$ is decomposed into contributions of the form of $r_8$ it is very useful to use twistor variables. If we take the twistors describing the two-dimensional ten-sided contour to be given by $\{W_1,\ldots,W_{10} \}$, then the twistors for the three eight-sided loops of Fig. \ref{10ptto8pt} are given by 
\begin{eqnarray}
&C_a &: \quad \{W_1,W_4,W_5,W_6,W_7,W_8,W_9,W_{10}\}\,, \notag \\
&C_b &: \quad \{W_1,W_2,W_3,W_6,W_7,W_8,W_9,W_{10}\}\,, \notag \\
&C_{ab} &: \quad \{W_1,W_4,W_3,W_6,W_7,W_8,W_9,W_{10}\}\,. 
\end{eqnarray}
The twistor configurations of the original loop and the three reduced loops are shown in Fig. \ref{twistor10ptto8pt}. We see that the points $x_a$ and $x_b$ correspond to the twistor lines $(1\ov{4})$ and $(3\ov{6})$ respectively.

Thus we have that $r_{10}$ is given by 
\begin{align}
r_{10} (1,\ov{2},3,\ov{4},5,\ov{6}|7,\ov{8},9,\ov{10}) = \,\,& r_8(1,\ov{4},5,\ov{6}|7,\ov{8},9,\ov{10})
 + r_8(1,\ov{2},3,\ov{6}|7,\ov{8},9,\ov{10}) \notag \\&- r_8(1,\ov{4},3,\ov{6}|7,\ov{8},9,\ov{10}) \notag \\
= \,\, & L(1,5,7,9)L(\ov{4},\ov{6},\ov{8},\ov{10}) +  L(1,3,7,9)L(\ov{2},\ov{6},\ov{8},\ov{10}) \notag \\& -  L(1,3,7,9)L(\ov{4},\ov{6},\ov{8},\ov{10})
\end{align}
For both $w$ and $\ov{w}$ variables this is exactly of the form of an invariant at five points constructed from four-point ones as in equation (\ref{5ptfrom4pt}).
We thus see that $r_{10}$ is invariant under two copies of the Yangian $Y(sl(2))$ with representation labels $\vec{\nu} = (1,1,1,-1,-1)$ and $\vec{\bar{\nu}} = (1,1,1,-1,-1)$.

Indeed quite generally we find that any definition of the reference square gives a Yangian invariant form for the function $r_n$. Let us take the general case in the two-dimensional kinematics with several kinks on the top and bottom sides of the reference square (see Fig \ref{2dpendiag}). As described in \cite{Gaiotto:2010fk} we can use the reduction argument step by step to reduce the number of kinks on each side until we are left with a sum over eight-sided contributions. 
Then the reduction procedure means we can write
\begin{align}
r_{n}(1,\bar{2},\ldots,\ov{i-1}|i,\ov{i+1},\ldots,\overline{n}) = & \,\,\phantom{+} r_{n-2}(1,\ov{4},5,\ov{6},\ldots|i,\ldots,\ov{n})\notag \\
& + r_{n-2}(1,\ov{2},3,\ov{6},\ldots|i,\ldots,\ov{n}) \notag \\
&- r_{n-2}(1,\ov{4},3,\ov{6},\ldots|i,\ldots,\ov{n})\,.
\end{align}
We can apply the same mechanism to the right group until we arrive at an expression made from many terms of the form of $r_8(1,\ov{p},q,\ov{i-1}|i,\ov{s},t,\ov{n})$ for different values of $p,q,s$ and $t$. All such terms are invariants of the two copies of the Yangian $Y(sl(2))$ with the representation labels $\vec{\nu}=(1,\ldots,1,-1,\ldots,-1)$ and similarly for $\vec{\bar{\nu}}$. Here the labels $\nu_i$ and $\ov{\nu}_i$ corresponding to twistors in the left group take the value $1$ while those for the right group take the value $-1$. Thus we conclude the $r_{n}$ is always invariant under the two commuting Yangians with a natural representation corresponding to the choice of OPE channel (i.e. choice of reference square).

The analysis we have presented here is the first indication that the Yangian symmetry seen at the level of the tree amplitudes (or the integrand for loop amplitudes) exhibits itself in a simple and natural way on the functions at one loop. Of course we have only two copies of the bosonic Yangian $Y(sl(2))$ in the Wilson loop problem, not the full $Y(psl(4|4))$, however we believe this is a firm indication that the symmetry is still present at loop level, acting in a predictive way and constraining the amplitudes. The fact that the conformal symmetry extends to its Yangian is the effect of the original conformal symmetry of the scattering amplitudes. At tree level the amplitudes can be recursively defined via the BCFW relations, with each term being invariant by itself under the Yangian symmetry.
The situation we have seen here for the Wilson loops at one loop is similar to that at tree-level for the scattering amplitudes. We find that the relevant finite part of the Wilson loop is defined recursively down to the octagonal loop. Each term appearing in the recursive procedure is invariant on its own under the same representation of both copies of the Yangian $Y(sl(2))$. The symmetry is expressed as certain second-order equations acting on the ratio $r_n$. In this respect it similar to the equations in \cite{Drummond:2010cz} acting on individual loop integrals of a certain type. One difference here is that the equations for $r_n$ are homogeneous while those for the loop integrals are inhomogeneous. Furthermore the differential equations for the loop integrals were valid in arbitrary kinematics while here we have restricted ourselves to the two-dimensional setup.
It will be very interesting to understand if the symmetry persists beyond one loop and whether it extends beyond the two-dimensional kinematics we have examined here. In this regard we should point out that the results we have found for the light-like Wilson loops hold in any gauge theory since we are only looking at one-loop expressions. Of course if the symmetry is to persist beyond one loop it would be most natural to find it in the planar limit of the $\mathcal{N}=4$ theory.

\section*{Acknowledgements}
We would like to thank Johannes Henn for discussions on the issues raised here and for collaboration on closely related topics.

\bibliography{invariants}
\bibliographystyle{nb}

\end{document}